\documentclass{kkzm22}

\usepackage[T1]{fontenc} 
\usepackage[cp1250]{inputenc}
\usepackage{algorithmic}
\usepackage{algorithm}
\usepackage{pdflscape}
\usepackage{graphicx}

\usepackage{enumitem}

\newtheorem{theorem}{Theorem}
\newtheorem{definition}{Definition}
\newtheorem{remark}{Remark}

\newcommand{\be}{\begin{equation}} \newcommand{\ee}{\end{equation}}
\newcommand{\bd}{\begin{displaymath}} \newcommand{\ed}{\end{displaymath}}
\newcommand{\ba}{\begin{align}} \newcommand{\ea}{\end{align}}
\newcommand{\baa}{\begin{align*}} \newcommand{\eaa}{\end{align*}}
\newcommand{\ben}{\begin{enumerate}} \newcommand{\een}{\end{enumerate}}
\newcommand{\bi}{\begin{itemize}} \newcommand{\ei}{\end{itemize}}

\newcommand{\ud}{\mathrm{d}}

\newcommand{\Expectation}[1]{\operatorname{E}\left[ #1 \right]}

\newcommand{\variance}[1]{\operatorname{Var}\left[ #1 \right]}

\title{Trait evolution with jumps: illusionary normality}
\author{Krzysztof Bartoszek$^
{1}$}

\skrocone{K.~Bartoszek}{Phylogenetic jumps}

\affiliation{
\text{
$^{1}$Department of Computer and Information Science, Link\"oping University, } \\
\text{Link\"oping 581 83, Sweden,} \\
\text{
$^{1}$\email{bartoszekkj@gmail.com}}\\
}

\begin{document}
\maketitle

\begin{abstract}
Phylogenetic comparative methods for real--valued traits usually make use of 
stochastic process whose trajectories are continuous.
This is despite biological intuition that evolution is rather punctuated than
gradual. On the other hand, there has been a number of recent proposals of evolutionary
models with jump components. However, as we are only beginning to understand
the behaviour of branching Ornstein--Uhlenbeck (OU) processes the asymptotics
of branching  OU processes with jumps is an even greater unknown. In this
work we build up on a previous study concerning OU with jumps evolution on a pure birth tree.
We introduce an extinction component and explore via simulations, its effects 
on the weak convergence of such a process.
We furthermore, also use this work to illustrate the simulation and graphic generation possibilities
of the mvSLOUCH package.
\end{abstract}

\section{Introduction}

Contemporary stochastic differential equation (SDE) models of 
continuous trait evolution are focused around the 
Ornstein--Uhlenbeck process \cites{THan1997}
\be
\label{eqSDEOU}
\ud X(t) = -\alpha(X(t)-\theta(t))\ud t + \sigma_{a}\ud B(t),
\ee
where $\theta(t)$ can be piecewise linear. These models take into account
the phylogenetic structure between the contemporary species.
The trait follows the SDE along each branch of the tree (with possibly branch specific parameters). 
At speciation times this process divides into two processes evolving independently from that point. 

However, the fossil record indicates \cites{NEldSGou1972,SGouNEld1977,SGouNEld1993} that change is not as gradual as Eq. \eqref{eqSDEOU} suggests. 
Rather, that jumps occur and that the framework of L\'evy processes could be more appropriate.
There has been some work in the direction of phylogenetic Laplace motion \cites{KBar2012,PDucCLeuSSziLHarJEasMSchDWeg2017,MLanJSchMLia2013}
and jumps at speciation points \cites{KBar2014,KBar2016arXiv,FBok2003,FBok2008}.
In this work we describe recent asymptotic results on such a jump model (developed in \cite{KBar2016arXiv})
and develop them by including an extinction component. 

It is worth pointing out
that OU with jumps (OUj) models are actually very attractive from a biological
point of view. They seem to capture a key idea behind the theory of punctuated
equilibrium (i.e. the theory of evolution with jumps \cite{SGouNEld1977}). 
At a branching event something dramatic (the jump) could have occurred that drove species apart.
But then ``The further removed in time a species from the original
speciation event that originated it, the more its genotype will have become stabilized and the more
it is likely to resist change.'' \cite{EMay1982}. Therefore, between branching events (and jumps) we could
expect stasis---``fluctuations of little or no accumulated consequence'' taking place
\cite{SGouNEld1993}. This fits well with an OUj model. If $\alpha$ is large enough, then
the process approaches its stationary distribution rapidly and the stationary oscillations around
the (constant) mean can be interpreted as stasis between jumps.

In applications the phylogeny is given (from molecular sequences) but when the aim
is to study large sample properties some model of growth has to be assumed.
A typical one is the constant rate birth--death process
%The phylogeny is modelled by a pure birth process with speciation rate $\lambda$
conditioned on $n$ contemporary tips (cf. \cites{TGer2008a,TGer2008b,SSagKBar2012}).
%The key property of this branching process is that the 
%time between speciation events $k$ and $k+1$ is exponential with
%parameter $k$. This is immediate from the memoryless
%property of the process and the distribution of the minimum
%of exponential random variables. This allows us to calculate 
%Laplace transforms of relevant speciation times \cite{KBar2014,KBar2016arXiv,KBarSSag2015a,SSagKBar2012}.

Regarding the phenotype 
the Yule--Ornstein--Uhlenbeck with jumps (YOUj) model was recently considered  \cite{KBar2016arXiv}.
The phylogeny is a pure birth process (no extinction) and 
the trait follows an OU process. However, just after the $k$--th 
branching point
($k=1,\ldots,n-1$, counting from the root with the root being the first branching point)
on the phylogeny, with a probability 
$p_{k}$, independently on each daughter lineage, a jump can occur. 
The jump is assumed to be normally distributed with mean $0$
and variance $\sigma_{c,k}^{2} < \infty$. 
Just after a speciation event at time $t$, independently for each daughter lineage, the trait
value $X(t)$ will be 
\be\label{eqProcJ}
X^{+}(t) = (1-Z)X(t^{-}) + Z(X(t^{-})+Y).
\ee
In the above Eq. \eqref{eqProcJ} $X(t^{-/+})$ means the value of $X(t)$ respectively just before and 
after time $t$, $Z$ is a binary random variable with probability $p_{k}$ of being 
$1$ (i.e. jump occurs) and \mbox{
$Y\sim \mathcal{N}(0,\sigma_{c,k}^{2})$.}

\section{The pure birth case}
In the case of the pure birth tree central limit theorems (CLTs) can be explicitly
derived thanks to a key property of the pure birth process.
%Assuming a speciation rate $\lambda=1$
The time between speciation events $k$ and $k+1$ is exponential with
parameter $\lambda k$ due to the memoryless
property of the process and the law of the minimum
of i.i.d. exponential random variables. This allows us to calculate 
Laplace transforms of relevant speciation times 
and count speciation events on lineages
\cites{KBar2014,KBar2016arXiv,KBarSSag2015a,SSagKBar2012}.

We first remind the reader about the mathematical concept of sequence
convergence with density $1$ (see e.g. \cite{KPet1983}) and then summarize the previously obtained CLTs.
\begin{definition}
A subset $E \subset \mathbb{N}$ of 
positive integers is said to have density $0$  if 
$$
\lim\limits_{n \to \infty}\frac{1}{n}\sum\limits_{k=0}^{n-1}\chi_{E}(k) =0,
$$
where $\chi_{E}(\cdot)$ is the indicator function of the set $E$.
\end{definition}
\begin{definition}
A sequence $a_{n}$ converges to $0$ with density $1$ if 
there exists a subset $E\subset \mathbb{N}$ of 
density $0$ such that 
$$
\lim\limits_{n \to \infty,n \notin E}a_{n} =0.
$$
\end{definition}
Let $\overline{Y}_{n} = (\overline{X}_{n}-\theta)/\sqrt{\sigma_{a}^{2}/2\alpha}$ be the normalized sample mean
of the YOUj process with $\overline{Y}_{0}=\delta$. 
Denote by $\mathcal{Y}_{n}$ the $\sigma$--algebra containing information on the tree and jump pattern
($\overline{Y}_{n}$ conditional on $\mathcal{Y}_{n}$ is normal). Assume also that $\lambda=1$.
The restriction
of $\lambda=1$ is a mild one, changing $\lambda$ is equivalent to rescaling
time.
\begin{theorem}[\cite{KBar2016arXiv}]\label{thmCLTYOUjpsConst}
Assume that the jump probabilities and jump variances are constant equalling $p$ and
$\sigma_{c}^{2}< \infty$ respectively.
%Let $\overline{Y}_{n} = (\overline{X}_{n}-\theta)/\sqrt{\gamma_{n}}$ be the normalized sample mean
%of the YOUj process with $\overline{Y}_{0}=\delta^{\ast}_{0}$. 
The process $\overline{Y}_{n}$ has the following, depending on $\alpha$, 
asymptotic with $n$ behaviour.
\begin{enumerate}[label=(\Roman*)]
\item \label{rCLTthmpIpConst} If $0.5<\alpha$, 
then the conditional variance of the scaled sample mean
$\sigma_{n}^{2}:=n\variance{\overline{Y}_{n} \vert \mathcal{Y}_{n}}$
%is a submartingale that 
converges in $\mathbb{P}$ to a random variable $\sigma_{\infty}^{2}$
with mean 
$$
\begin{array}{rcl}
\Expectation{\sigma_{\infty}^{2}} & = & 1+\frac{2p\sigma^{2}_{c}}{\sigma_{a}^{2}}+ \frac{4p\sigma^{2}_{c}}{2\alpha(2\alpha-1)\sigma_{a}^{2}}.
\end{array}
$$
The scaled sample mean,
$\sqrt{(n)}~\overline{Y}_{n}$ converges weakly to a random variable
%whose characteristic function can be expressed as
%whose moment generating function can be expressed as
whose characteristic function can be expressed in terms of the Laplace transform of 
$\sigma_{\infty}^{2}$
$$
%M_{\sqrt{(n)}~\overline{Y}_{n}}(x) = M_{\sigma_{\infty}^{2}}(x^{2}/2).
%\phi_{\sqrt{(n)}~\overline{Y}_{n}}(x) = \phi_{\sigma_{\infty}^{2}}(ix^{2}/2).
\phi_{\sqrt{(n)}~\overline{Y}_{n}}(x) = \mathcal{L}(\sigma_{\infty}^{2})(x^{2}/2).
%\Expectation{e^{-\sigma_{\infty}^{2}x^{2}/2}}..
$$
\item \label{rCLTthmpIIpConst} If $0.5=\alpha$, 
then the conditional variance of the scaled sample mean
$\sigma_{n}^{2}:=n\ln^{-1} n\variance{\overline{Y}_{n} \vert \mathcal{Y}_{n}}$
%is a submartingale that 
converges in $\mathbb{P}$
to a random variable $\sigma_{\infty}^{2}$
with mean 
$$
\begin{array}{rcl}
\Expectation{\sigma_{\infty}^{2}} & = & 2 + 4p\sigma_{c}^{2}(\sigma_{a}^{2}/(2\alpha))^{-1}.
\end{array}
$$
The scaled sample mean,
$\sqrt{(n/\ln n)}~\overline{Y}_{n}$ converges weakly to a random variable
%whose moment generating function can be expressed as
%whose characteristic function can be expressed as
whose characteristic function can be expressed in terms of the Laplace transform of 
$\sigma_{\infty}^{2}$
$$
\phi_{\sqrt{(n/\ln n)}~\overline{Y}_{n}}(x) = \mathcal{L}(\sigma_{\infty}^{2})(x^{2}/2).
%\phi_{\sqrt{(n/\ln n)}~\overline{Y}_{n}}(x) = \phi_{\sigma_{\infty}^{2}}(ix^{2}/2).
%M_{\sqrt{(n/\ln n)}~\overline{Y}_{n}}(x) = M_{\sigma_{\infty}^{2}}(x^{2}/2).
$$
\item \label{rCLTthmpIIIpConst} If $0<\alpha <0.5$, then %we can allow $\kappa_{n}$ to take any value in $[0,1]$ and 
$n^{\alpha}\overline{Y}_{n}$ converges almost surely and in $L^{2}$ to a random variable
$Y_{\alpha,\delta}$ with first two moments
$$
\begin{array}{rcl}
\Expectation{Y_{\alpha,\delta}} & = & \delta \Gamma(1+\alpha), \\
\Expectation{Y_{\alpha,\delta}^{2}} & = & 
- (1-\delta^{2})\Gamma(2\alpha+1)
+\frac{1+2\alpha}{1-2\alpha}\Gamma(1+2\alpha)(1+2p\sigma_{c}^{2}(\sigma_{a}^{2}/(2\alpha))^{-1})
%\left(2p\sigma_{c}^{2}(\sigma_{a}^{2}/(2\alpha))^{-1}\frac{1+2\alpha}{1-2\alpha}
%- (1-\delta^{2})\right)\Gamma(1+2\alpha).
%\left(\delta^{2}+(1-\kappa_{\infty})\frac{4\alpha}{1-2\alpha}\right)\Gamma(1+2\alpha).
\end{array}
$$
\end{enumerate}
\end{theorem}
\begin{theorem}[\cite{KBar2016arXiv}]\label{thmCLTYOUjpsae0}
%Let $\overline{Y}_{n} = (\overline{X}_{n}-\theta)/\sqrt{\gamma_{n}}$ be the normalized sample mean
%of the YOUj process with $\overline{Y}_{0}=\delta^{\ast}_{0}$. 
%Let $\overline{Y}_{n} = (\overline{X}_{n}-\theta)/\sqrt{\sigma_{a}^{2}/2\alpha}$ be the normalized sample mean
%of the YOUj process with $\overline{Y}_{0}=\delta_{0}$. 
If $\sigma_{c,k}^{4}p_{k}$ is bounded and goes to $0$ with density $1$, 
then depending on $\alpha$ the process $\overline{Y}_{n}$ has the following 
asymptotic with $n$ behaviour. % depending on $\alpha$.
\begin{enumerate}[label=(\Roman*)]
\item \label{rCLTthmpI} If $0.5<\alpha$, 
%$\sigma_{c,n}^{2}\kappa_{n} \to 0$
then $\sqrt{(n)}~\overline{Y}_{n}$ is asymptotically normally distributed with mean $0$ and
variance $(2\alpha+1)/(2\alpha-1)$.
\item \label{rCLTthmpII} If $0.5=\alpha$, 
%and $\sigma_{c,n}^{2}\kappa_{n} \to 0$
then $\sqrt{(n/\ln n)}~\overline{Y}_{n}$ is asymptotically normally distributed with mean $0$ and
variance $2$.
\end{enumerate}
\end{theorem}
\begin{remark}
Notice that Thm. \ref{thmCLTYOUjpsae0} immediately implies the CLTs when there are no
jumps, i.e. $p_{k}=0$ for all $k$ \cite{KBarSSag2015a}.
\end{remark}
\section{Extinction present}
\begin{figure}[!ht]
\centering
\includegraphics[width=0.45\textwidth]{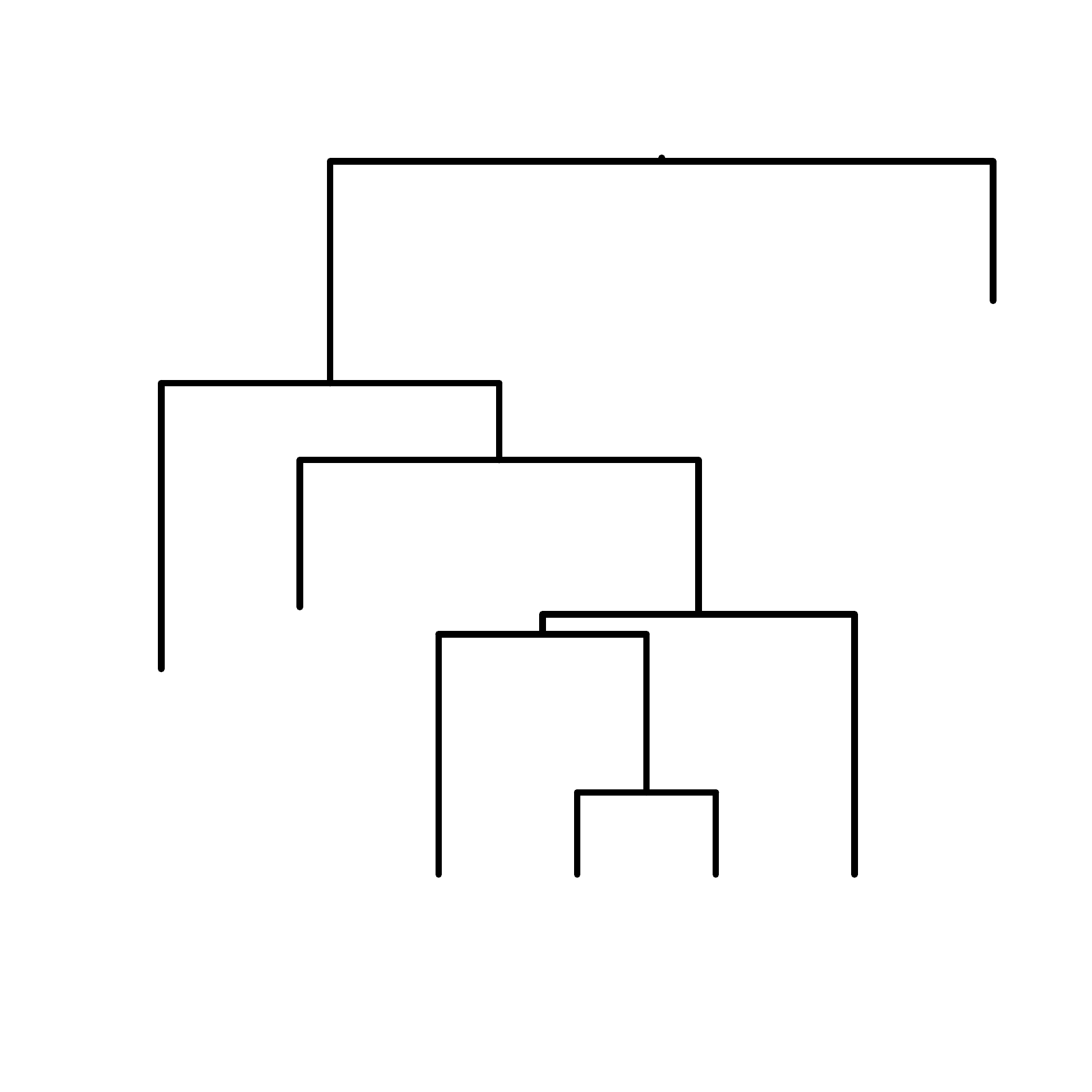}  
\includegraphics[width=0.45\textwidth]{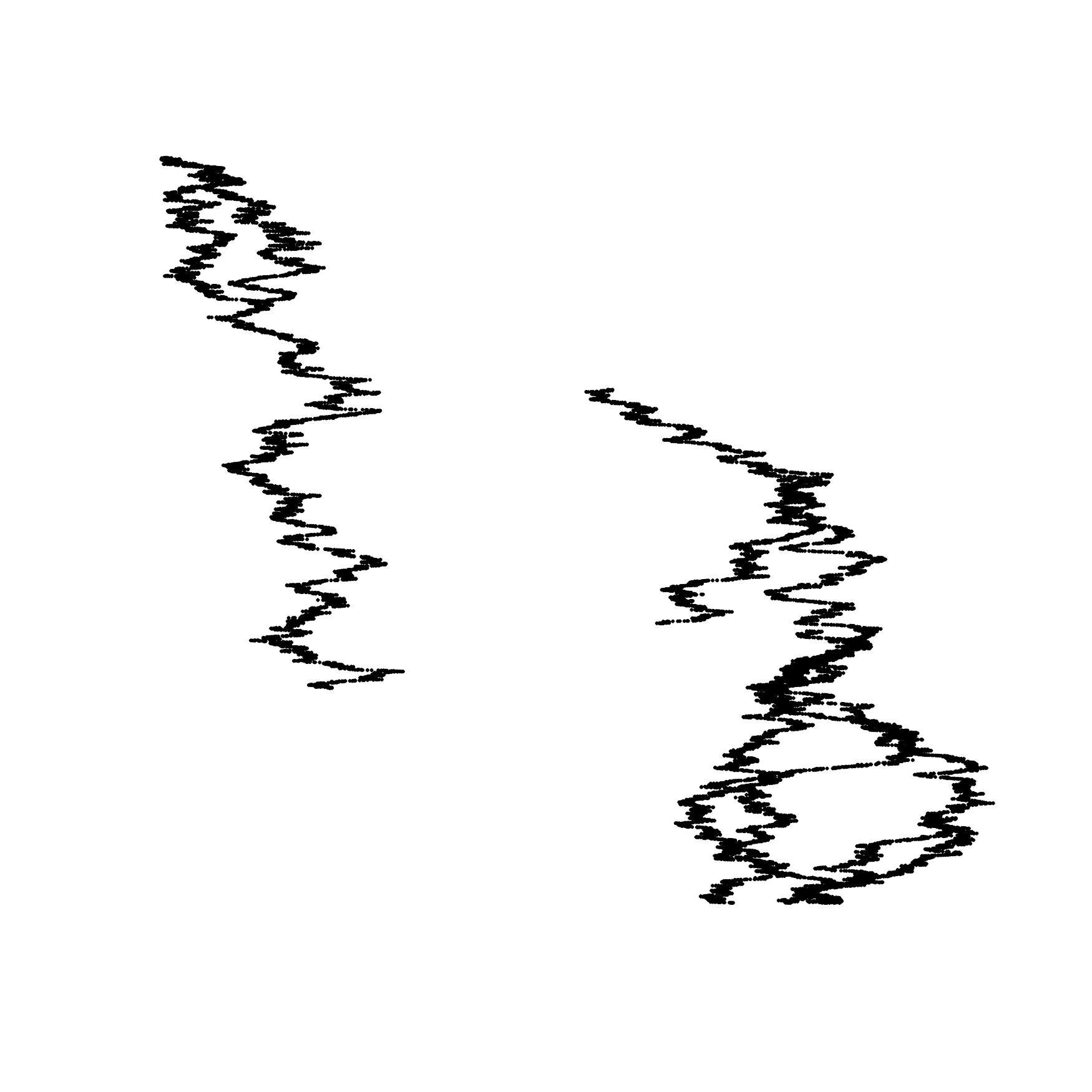} 
\caption{Left: birth--death tree with a clade of $4$ contemporary species. Right: OUj process evolving on this tree
(graphic by mvSLOUCH).
There is a single jump in the trait process at the first speciation event after the root. The OU process
is a slowly adapting one with large jump variance ($\alpha=0.3$, $\sigma_{a}^{2}=\sigma_{c,k}^{2}=4$, tree height: $2.427$).
}\label{figTree}
\end{figure}

The no extinction assumption is difficult to defend biologically, unless
one considers extremely young clades. Therefore, it is desirable
to generalize Thms. \ref{thmCLTYOUjpsConst} and \ref{thmCLTYOUjpsae0} to the
case when the extinction rate, $\mu$, is non--zero. However, there
are a number of intrinsic difficulties associated with such a generalization. 
We do not have the Laplace transforms of the time to coalescent 
of a random pair of tips (its expectation seems involved enough, \cite{SSagKBar2012}).
More importantly we seem to be unable to say much about the number of hidden speciation
events on a random lineage. We have to remember that a jump can be due to any speciation
event, including those that lead to extinct lineages. A lineage that survived till
today can have multiple branches stemming from it that faded away in the past, see Fig. \ref{figTree}. 

Furthermore, for the OU model of trait evolution, we need to know the distribution
(or Laplace transform) of the times between the speciation events. Without extinction, $\mu=0$,
this was simple. The time between speciation events $k$ and $k+1$ was exponential with rate $\lambda k$,
as the minimum of $k$ independent rate $\lambda$ exponential random variables. However,
when $\mu > 0$ we not only need to know the number of hidden speciation events between two 
non--hidden (i.e. leading to contemporary tips) speciation events but also the law of the time
between the hidden speciation. We do not know this law and furthermore as we are conditioning
on $n$ it is not entirely clear if times between speciation events will 
be independent (like they are in the pure birth case). 

Our question is whether we can expect 
counterparts of Thms. \ref{thmCLTYOUjpsConst} and \ref{thmCLTYOUjpsae0} to hold
when extinction is present, i.e. $0< \mu \le 1 = \lambda$.  
We are not aware of analytical results on the issues raised in the previous paragraph.
Hence, we will approach answering the problem by simulations.
Based on the results in \cite{RAdaPMil2015} we should 
expect a phase transition to occur for $1-\mu=\alpha/2$ (remember %that the birth--rate 
$\lambda=1$).

We simulate $1000$ birth--death trees for $\mu=0.25,0.5,0.75$ 
conditioned on $n=500$ contemporary tip species with the TreeSim
\cites{TreeSim1,TreeSim2} R package.
On each phylogeny we simulate an OU process with
$\alpha=(1-\mu)$, $(1-\mu)/2$, $(1-\mu)/4$ using the 
mvSLOUCH R package \cite{KBarJPiePMosSAndTHan2012}. 
In all simulations $\sigma_{a}^{2}=\sigma_{c,k}^{2}=1$,
$p_{k}=0.5$ (for all internal nodes) and $X_{0}=\theta=0$. 
For a given phylogeny, OU simulation pair we calculate the 
scaled sample average, 
\begin{equation}\label{eqbarYnJ} 
\overline{V}_{n}= (\sqrt{n})\overline{X}_{n}/\sqrt{\sigma_{a}^{2}/(2\alpha)}.
\end{equation}
We report
the results of the simulations by plotting histograms with a 
mean $0$ and variance equalling the scaled sample variance  normal curve.
\begin{figure}[!ht]
\centering
\includegraphics[width=0.32\textwidth]{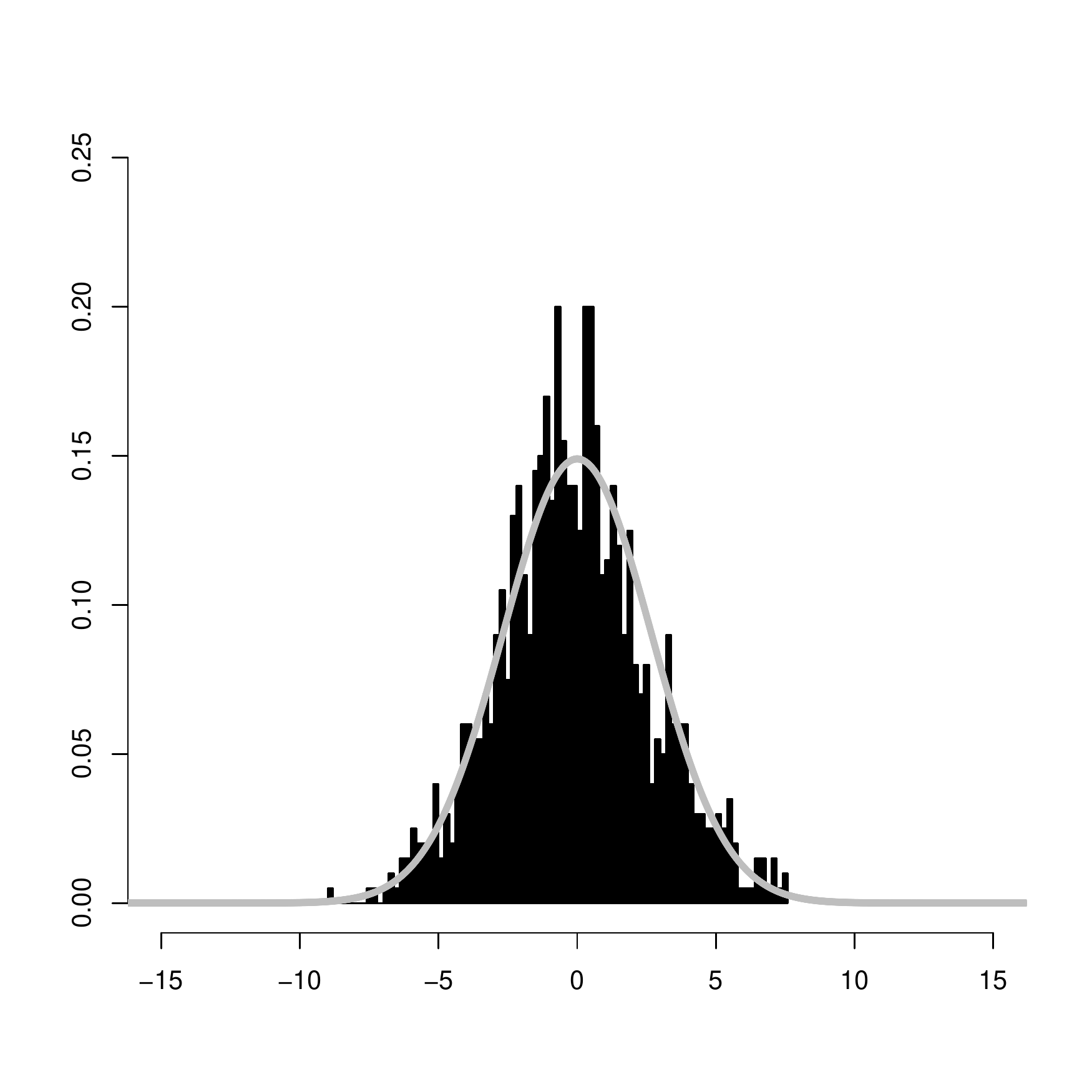}  
\includegraphics[width=0.32\textwidth]{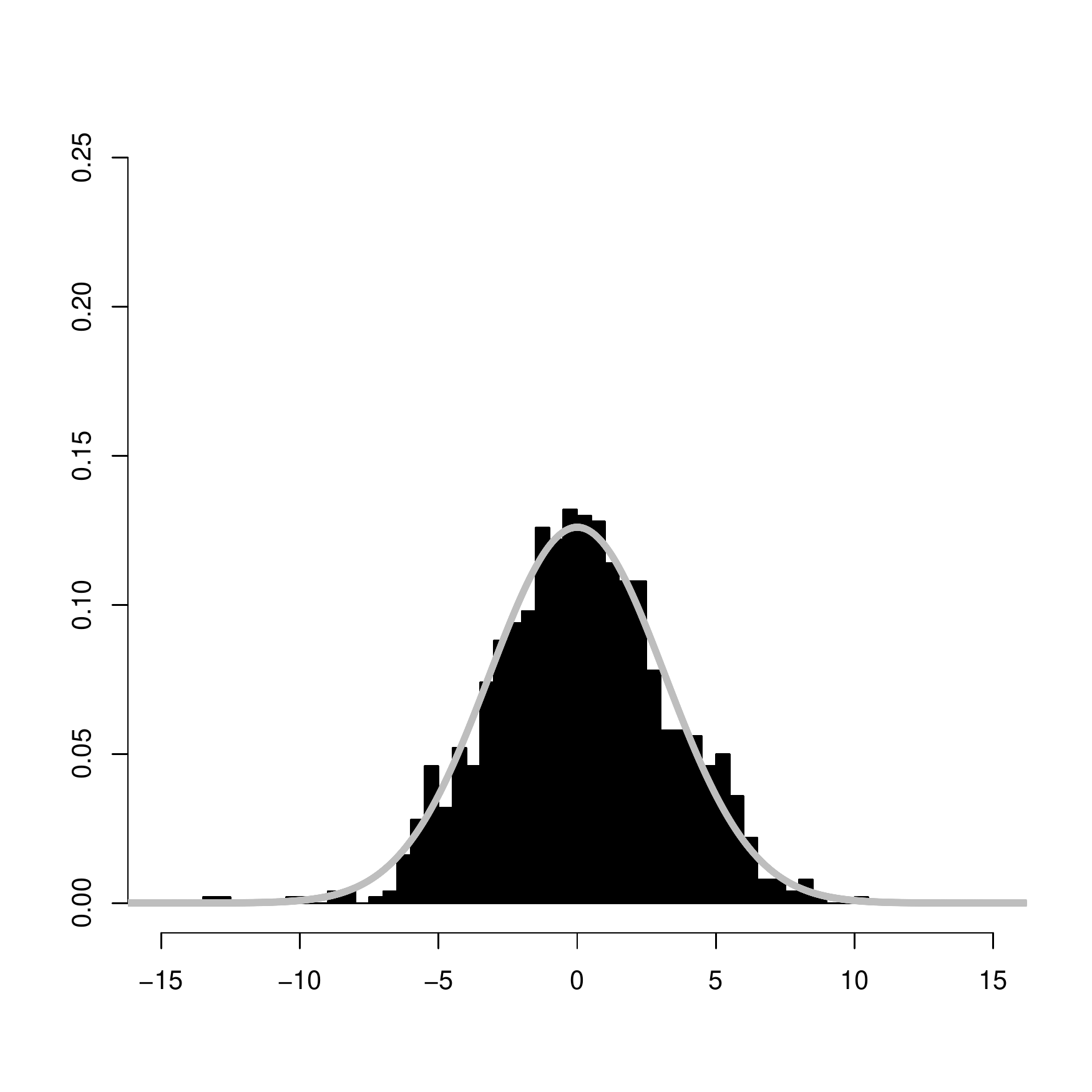}  
\includegraphics[width=0.32\textwidth]{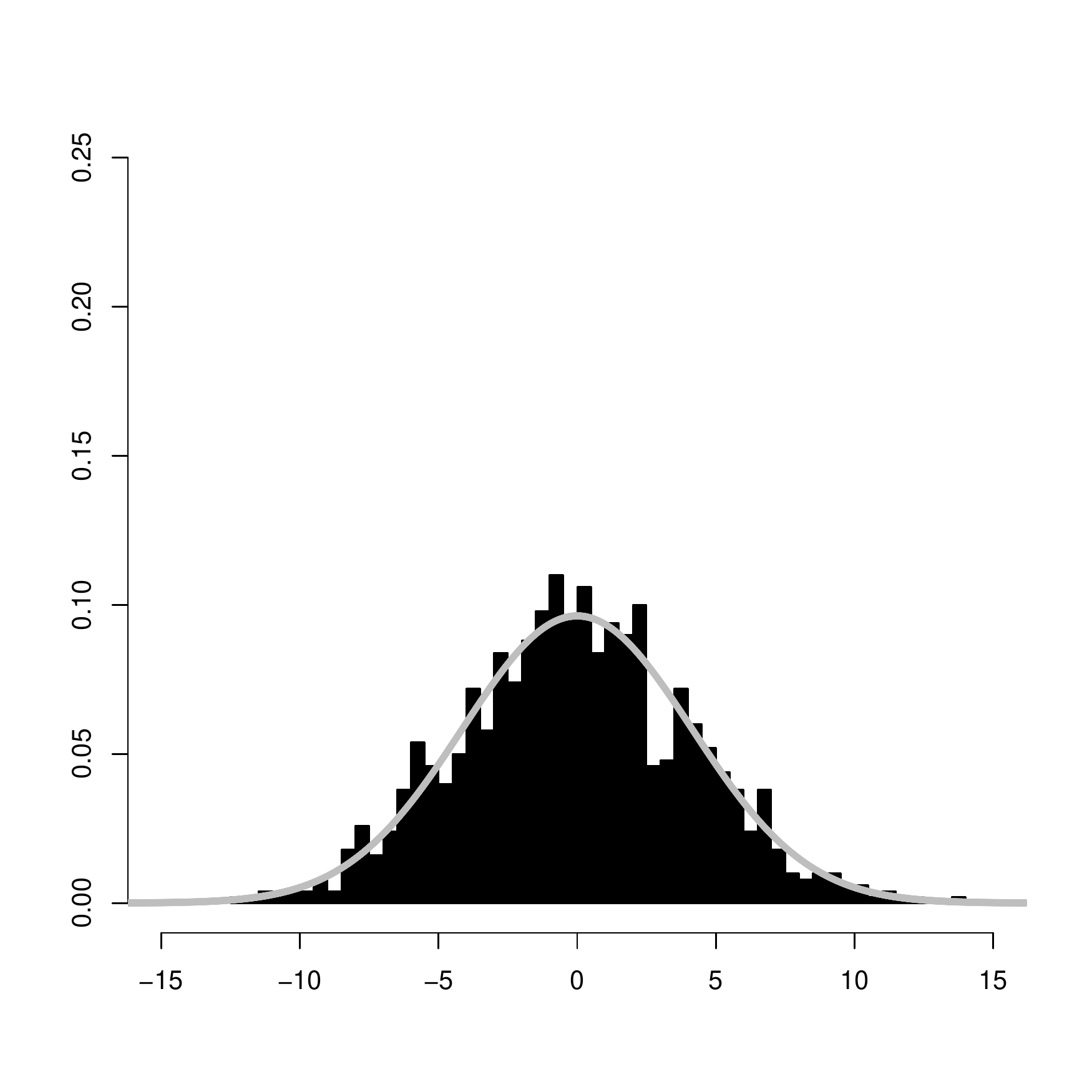} \\  
\includegraphics[width=0.32\textwidth]{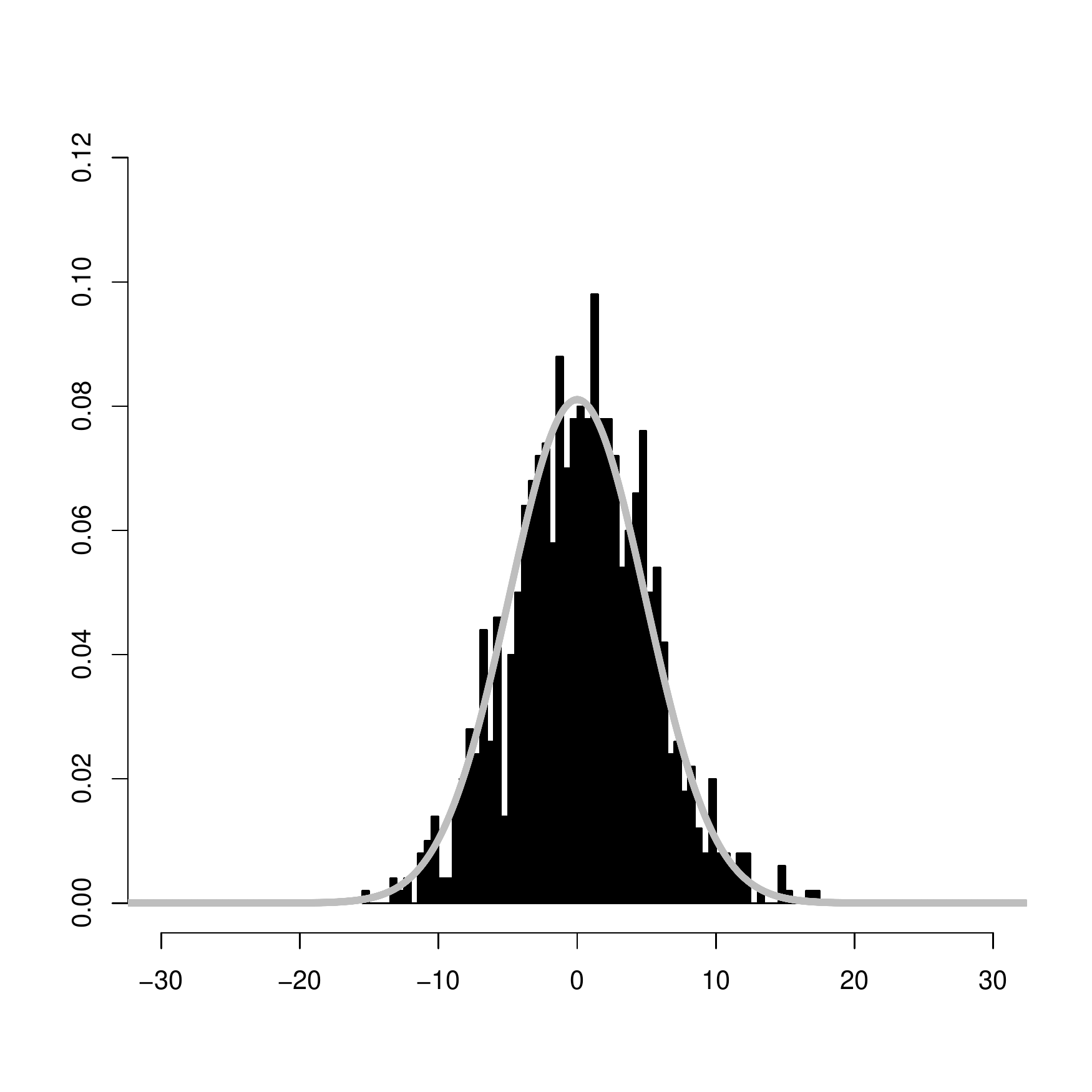}  
\includegraphics[width=0.32\textwidth]{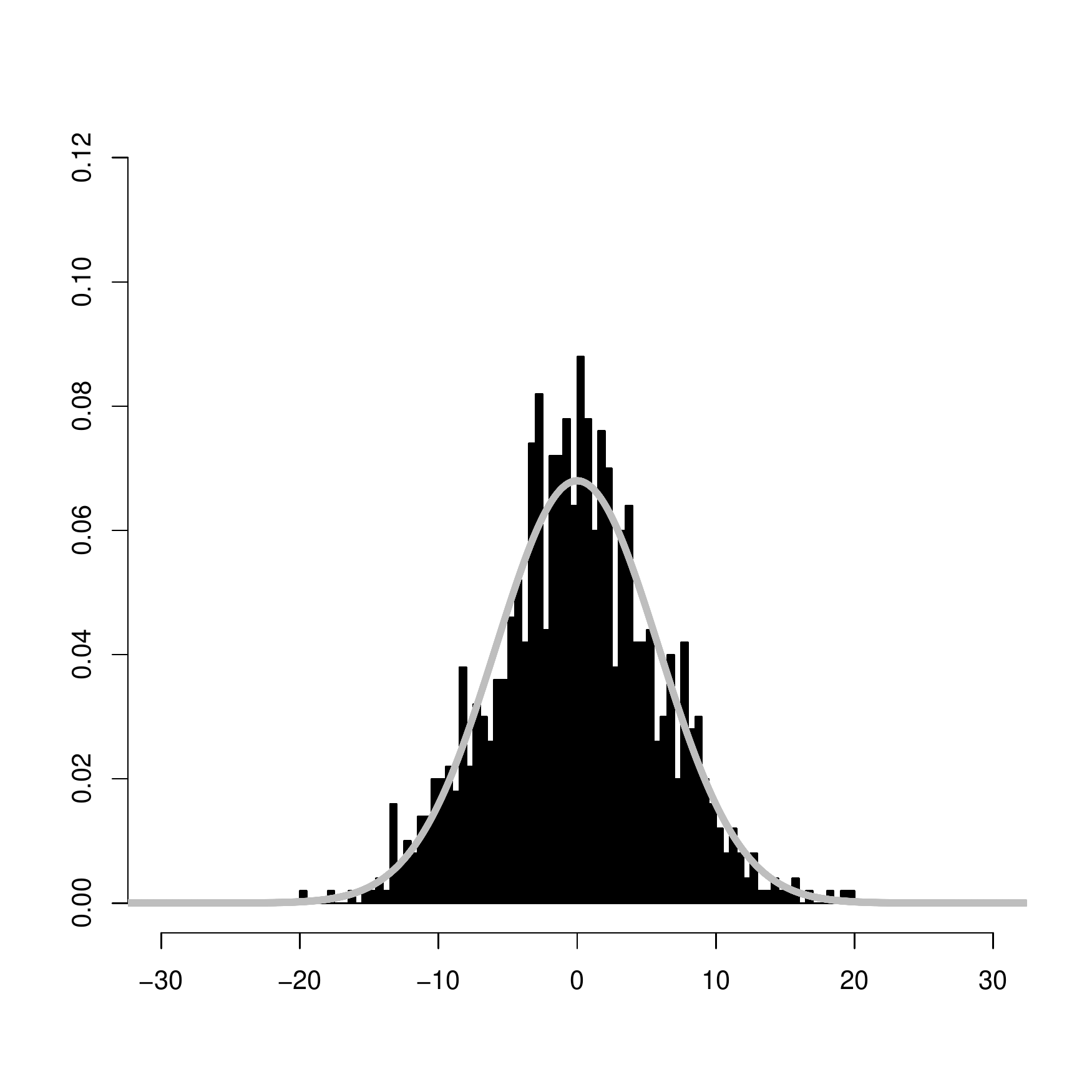}  
\includegraphics[width=0.32\textwidth]{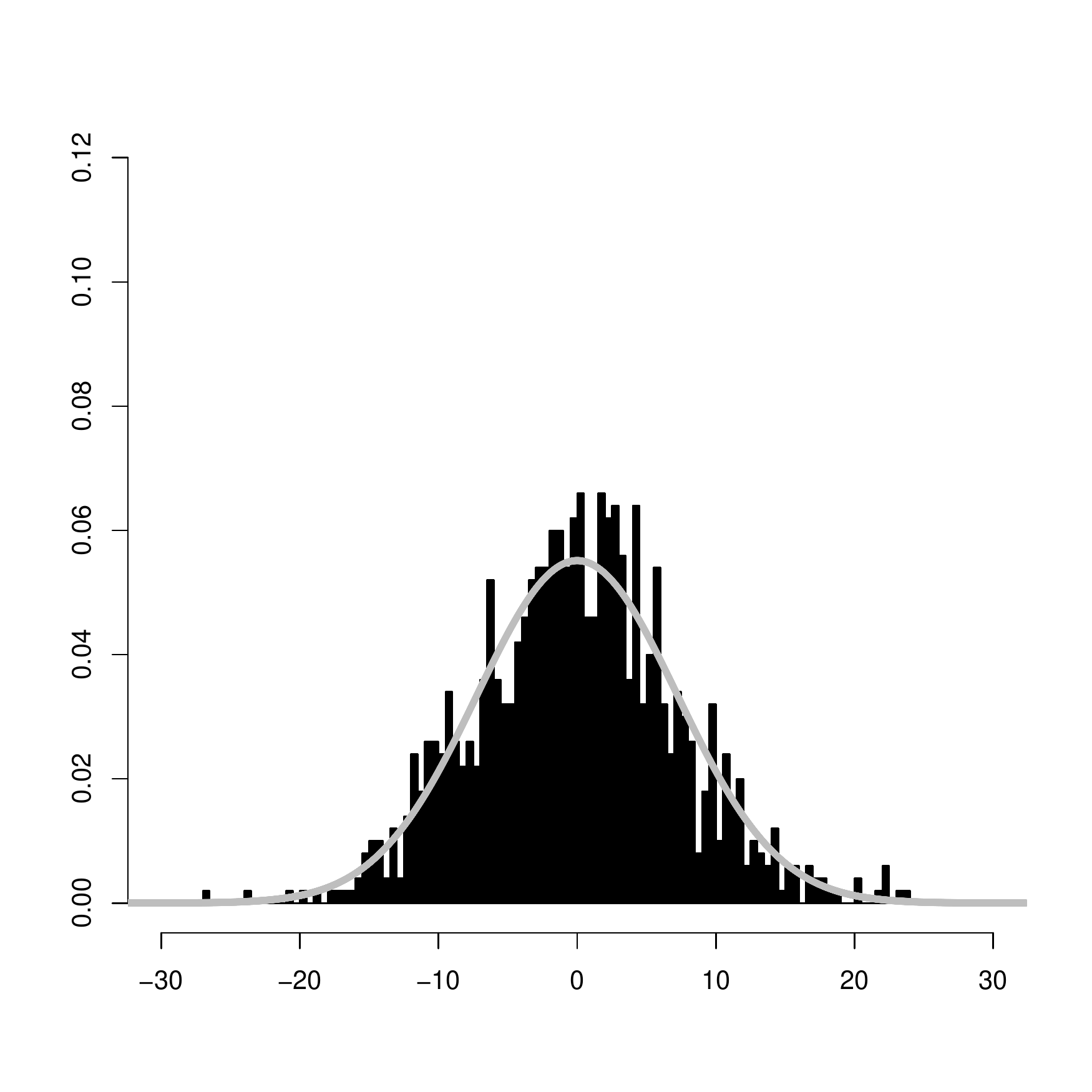} \\
\includegraphics[width=0.32\textwidth]{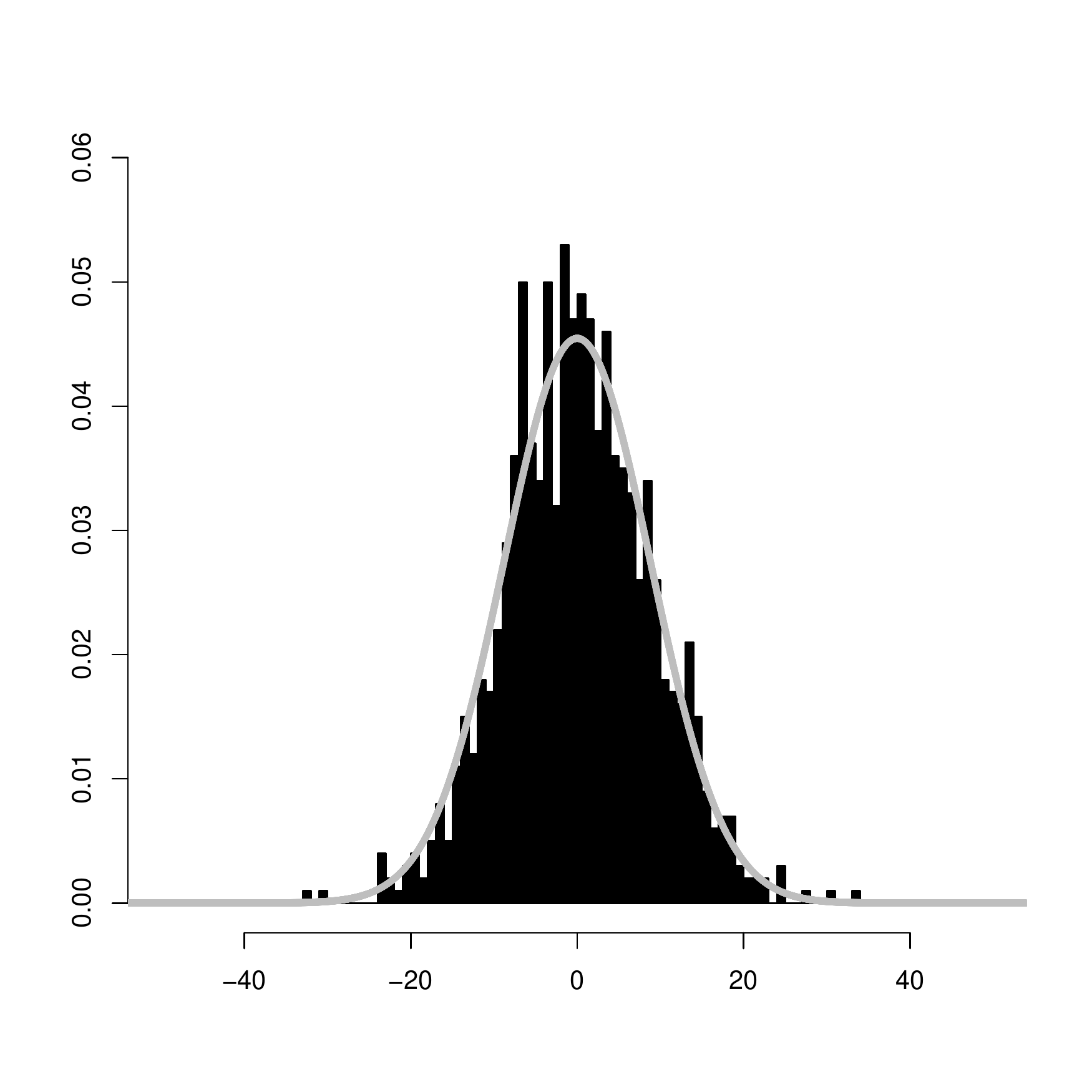}
\includegraphics[width=0.32\textwidth]{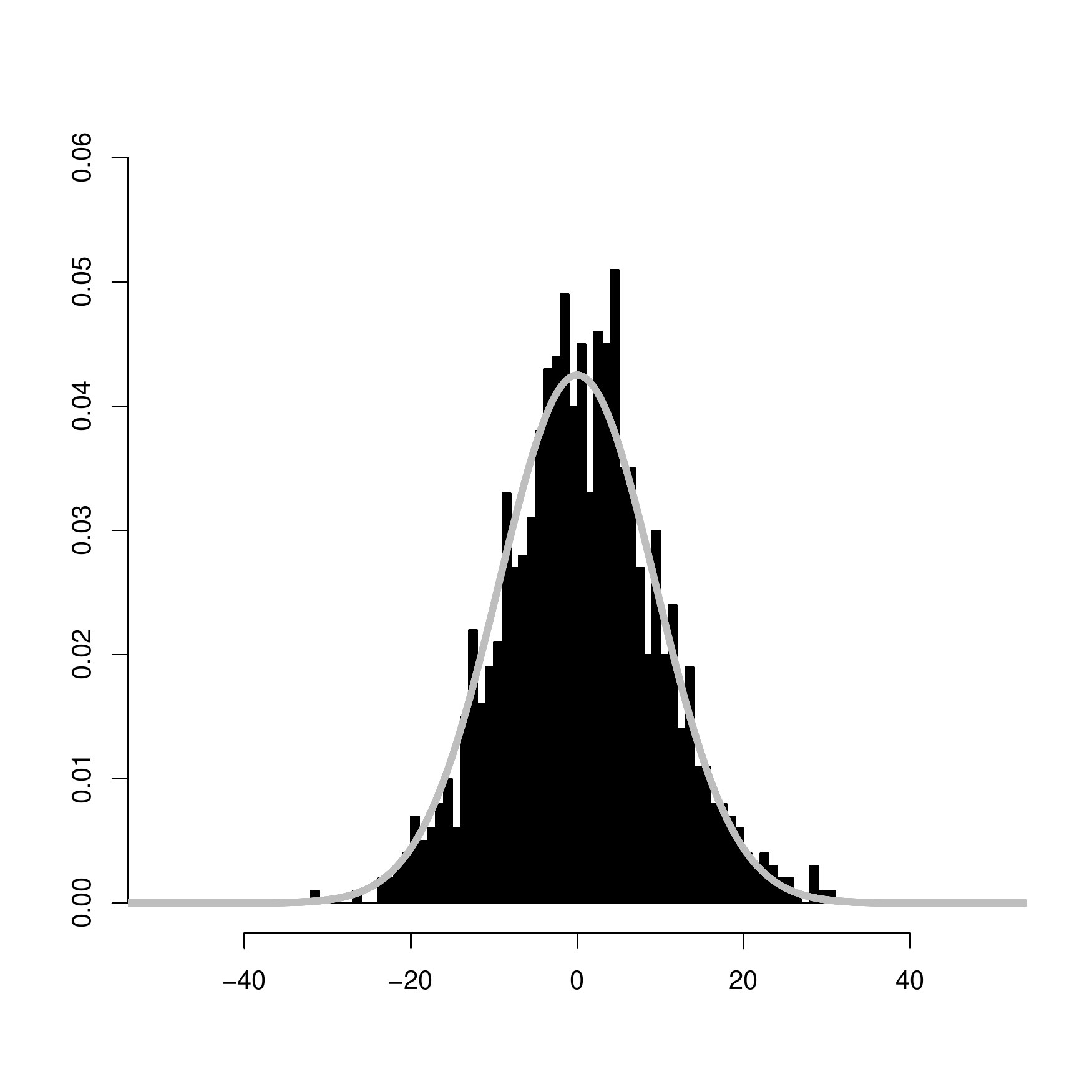}
\includegraphics[width=0.32\textwidth]{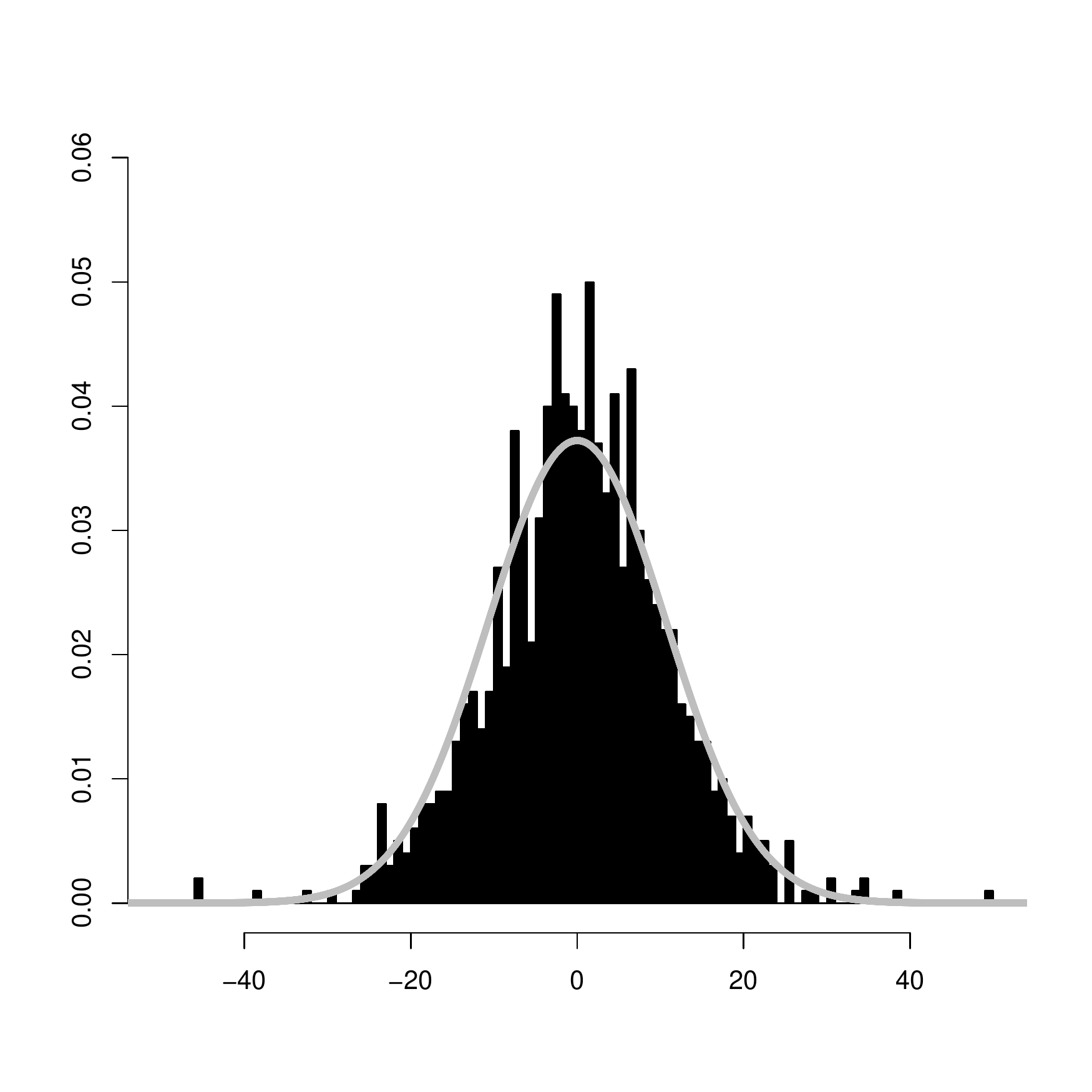}
%[]{}
%\includegraphics[]{}
\caption{Simulations of Eq. \eqref{eqbarYnJ}.
Left column $\mu=0.25$, centre column $\mu=0.5$, right column $\mu=0.75$,
top row $\alpha=(1-\mu)$, centre row $\alpha=(1-\mu)/2$ and bottom row $\alpha=(1-\mu)/4$.
The gray curve is the density curve of the normal distribution with mean $0$ and
variance equalling the sample variance. Each histogram is constructed from $1000$ simulated 
birth--death trees (birth rate $1$, death rate $\mu$) with an OUj process ($\theta=X_{0}=0$, $\sigma_{a}^{2}=\sigma_{c,k}^{2}=1$, $p_{k}=0.5$) 
evolving on top of the tree. Notice that the $x$ and $y$ axes differ between rows.
}\label{figBDSims}
\end{figure}
\begin{table}[!ht]
\begin{center}
\caption{Summary of simulated samples. The mean, variance, skewness and excess kurtosis refer to the sample moments of 
$\overline{V}_{n}$. The $95\%$ bootstrap confidence intervals for the excess kurtosis are 
based on $10000$ bootstrap replicates (of the $\overline{V}_{n}$ values) and calculated by the R package boot (basic bootstrap 
confidence intervals are reported).
%provided by the mrt R package \cite{DWriJHer2011}.
} \label{tabBDsims}
\begin{tabular}{|c|c|c|c|c|c|cl|}
\hline $\mu$ & $\alpha$ & $\alpha/(1-\mu)$ &  mean &  variance & skewness & \multicolumn{2}{c|}{excess kurtosis} \\
\hline $0.25$ & $0.75$ & $1$ & $-0.117$ & $7.173$ & $0.114$ & $0.046$ &$(-0.195,  0.27)$ \\
\hline $0.5$ & $0.5$ & $1$ & $0.116$ & $10.016$ & $-0.084$ & $0.280$ &$(-0.297,  0.803)$ \\
\hline $0.75$ & $0.25$ & $1$ & $-0.082$ & $17.158$ & $0.058$ & $-0.058$ &$(-0.318,  0.186)$ \\
\hline $0.25$ & $0.375$ & $0.5$ & $0.344$ & $24.227$ & $0.038$ & $0.110$& $(-0.185,  0.392)$ \\
\hline $0.5$ & $0.25$ & $0.5$ & $-0.141$ & $34.432$ & $0.012$ & $0.121$ &$(-0.186,  0.412)$ \\
\hline $0.75$ & $0.125$ & $0.5$ & $-0.056$ & $52.357$ & $0.087$ & $0.3$ &$(-0.064,  0.646)$ \\
\hline $0.25$ & $0.1875$ & $0.25$ & $0.212$ & $77.016$ & $0.072$ & $0.316$& $(-0.107,  0.722)$ \\
\hline $0.5$ & $0.125$ & $0.25$ & $0.575$ & $88.128$ & $0.102$ & $0.149$ &$(-0.14,  0.416)$ \\
\hline $0.75$ & $0.0625$ & $0.25$ & $0.3$ & $114.806$ & $-0.024$ & $1.169$ &$(0.328,  1.964)$ \\
\hline
\end{tabular}
\end{center}
\end{table}
\section{Illusionary normality?}
We report the result of our simulations in Fig. \ref{figBDSims} and Tab. \ref{tabBDsims}.
The histograms are not conclusive but they can be interpreted as indicating 
a similar as in Thms. \ref{thmCLTYOUjpsConst} and \ref{thmCLTYOUjpsae0} phase transition.
In the fast adaptation case, $\alpha=(1-\mu)>(1-\mu)/2$, the histograms do not seem
to deviate much from the normal curve. On the other hand, it is a bit surprising that
the ``worst'' looking histogram is when $\mu=0.25$, the one we would expect 
to be closest to ``normality''. As the ratio of $\alpha$ to $1-\mu$ decreases
we can see that the histograms of $\overline{V}_{n}$ deviate more from the normal
curve. Furthermore, when $\alpha=(1-\mu)/4<(1-\mu)/2$ we can start to see
heavier than normal tails. 

The analysis of the first four moments in Tab. \ref{tabBDsims} does point to 
two things. Firstly, the scaling by $(\sqrt{n})/(\sqrt{\sigma_{a}^{2}/(2\alpha))})$
is incorrect (cf. Thm. $3.3$ \cite{RAdaPMil2015}). In all setups the sample variance of $\overline{V}_{n}$ 
is much greater than $1$. However, based on the estimates of skewness and excess kurtosis we 
cannot reject normality outright. Only in the most extreme case ($\mu=0.75$, $\alpha/(1-\mu)=0.25$)
do the $95\%$ bootstrap confidence intervals of the excess kurtosis not cover $0$.
When $\alpha=(1-\mu)/2$ we should actually expect to have a logarithmic correction in the scaling,
$\sqrt{n\ln n}$, however we present here the histograms without it. Such
a logarithmic correction did not bring in any qualitative changes to the results 
and hence, for easiness of comparison in Tab. \ref{tabBDsims} and Fig. \ref{figBDSims}
we refrain from using it.

Based on the histograms and analysis of the first four moments we would not suspect
that with a constant jump probability $p$ we do not have a nearly classical CLT, i.e. 
weak convergence to a normal limit after scaling by $\sqrt{(n)}$. All that we would
think would remain, would be to find the correct variance of the limit. In fact
if we look at Fig. $2$ in \cite{KBar2016arXiv} we would be under a similar illusion.
The histogram for $\alpha=1$, $\lambda=1$, $\mu=0$, $p_{k}=0.5$ and $\sigma_{c,k}^{2}=1$
nearly perfectly fits into a normal density curve. However, Thms. \ref{thmCLTYOUjpsConst} 
and \ref{thmCLTYOUjpsae0} are very clear that for a constant product of 
$p_{k}\sigma_{c,k}^{4}$ we will not have a normal limit. 
Therefore, with $\mu>0$ we cannot expect a change of the situation. On the one
hand a non--zero extinction rate does cause the (conditional on $n$ contemporary tips)
tree to be higher. But on the other hand, with greater height comes to opportunity
for more jumps. And this variability of jump occurrences
seems to be the force pushing the limit away from normality.

The simulations and results presented here and in \cite{KBar2016arXiv} should also serve as a warning. 
Visual inspection of histograms from simulated data 
is never sufficient for drawing conclusions about a model's underlying
distribution. A bare minimum are goodness--of--fit tests but their conclusions should be supported by
analytical derivations.
\section{Acknowledgements}
{\small
%We would like to acknowledge Wojciech Bartoszek for many helpful comments and insights.
KB's research was supported by the Knut and Alice Wallenberg Foundation. 
KB's conference participation was supported by the Wenner--Gren Foundation (grant nr. RSh$2016$--$0078$).
%and the National Science Centre (grant nr. $2015\backslash 18\backslash \mathrm{E}\backslash \mathrm{NZ}8\backslash 00716$).

% Centre for Theoretical Biology at the University of Gothenburg,
%Svenska Institutets \"Ostersj\"osamarbete schoralship nr. 00507/2012,
%Stiftelsen f\"or Vetenskaplig Forskning och Utbildning i Matematik
%(Foundation for Scientific Research and Education in Mathematics),
%Knut and Alice Wallenbergs travel fund, Paul and Marie Berghaus fund, the Royal Swedish Academy of Sciences,
%and Wilhelm and Martina Lundgrens research fund.
}

%\textbf{ZROB BIBLIOGRSFIE}
%\bibliographystyle{plain}
%\bibliography{OUjCLT}
 \begin{bibliography}
% @preamble{ " \newcommand{\noop}[1]{} " }
\newcommand{\noop}[1]{}
 \vspace*{-0.3cm}
 
 \bib{RAdaPMil2015}{article}{
    author = {R. Adamczak},
    author={P. {Mi\l o\'s}},
    title = {{CLT} for {O}rnstein--{U}hlenbeck branching particle system},
    journal = {Elect. J. Probab.},
        volume={20},
        number={42},
        pages={1-35},
    year = {2015},
}

 \bib{KBar2012}{inproceedings}{
    TITLE={The {L}aplace Motion in Phylogenetic Comparative Methods},
    AUTHOR={K. Bartoszek},
    BOOKTITLE={Proceedings of the Eighteenth National Conference on Applications of Mathematics in Biology and Medicine, Krynica Morska},
        PAGES={25--30},
    YEAR={2012},    
}

\bib{KBar2014}{article}{
    TITLE={Quantifying the effects of anagenetic and cladogenetic evolution},
    AUTHOR={K. Bartoszek},
    JOURNAL={Math. Biosci.},
        VOLUME={254},
        PAGES={42-57},
    YEAR={2014},
}

\bib{KBar2016arXiv}{article}{
    TITLE={A Central Limit Theorem for punctuated equilibrium},
    AUTHOR={K. Bartoszek},
    journal = {ArXiv e-prints},
    %archivePrefix = "arXiv",
    pages = {1602.05189},
    YEAR={2016},
}

\bib{KBarJPiePMosSAndTHan2012}{article}{
    AUTHOR={K. Bartoszek},
    author={J. Pienaar},
    author={P. Mostad},
    author={S. Andersson},
    author={T. F. Hansen},
    TITLE={A phylogenetic comparative method for studying multivariate adaptation},
    JOURNAL={J. Theor. Biol.},
        VOLUME={314},
        PAGES={204-215},
    YEAR={2012},
}

\bib{KBarSSag2015a}{article}{
   author = {K. Bartoszek},
   author={S. Sagitov},
    title = {Phylogenetic confidence intervals for the optimal trait value},
  journal = {J. App. Prob.},
        volume ={52},
    pages={1115-1132},
     year = {2015},
}

\bib{FBok2003}{article}{
    AUTHOR={F. Bokma},
    TITLE={Testing for equal rates of cladogenesis in diverse taxa},
    JOURNAL={Evolution},
        VOLUME={57},
        NUMBER={11},
        PAGES={2469-2474},
    YEAR={2003},
}

\bib{FBok2008}{article}{
    AUTHOR={F. Bokma},
    TITLE={Detection of ``punctuated equilibrium'' by {B}ayesian estimation of speciation and extinction rates, ancestral character states, and rates of anagenetic and cladogenetic evolution on a molecular phylogeny},
    JOURNAL={Evolution},
        VOLUME={62},
        NUMBER={11},
        PAGES={2718-2726},
    YEAR={2008},
}

\bib{PDucCLeuSSziLHarJEasMSchDWeg2017}{article}{
  Author         = {P. Duchen},
  author={C. Leuenberger},
  author={S. M. Szil{\`a}gyi},
  author={L. Harmon},
  author={J. Eastman},
  author={M. Schweizer},
  author={D. Wegmann},
  Title          = {Inference of evolutionary jumps in large phylogenies using {L}{\'e}vy processes},
  Journal        = {Syst. Biol.},
  year           = {\noop{3001}in press 2017},
}

\bib{NEldSGou1972}{incollection}{
    author={N. Eldredge},
    AUTHOR={S. J. Gould},
    TITLE={Punctuated equilibria: an alternative to phyletic gradualism},
    BOOKTITLE={Models in Paleobiology},
    EDITOR={T. J. M. Schopf}, 
    EDITOR={J. M. Thomas},
    PUBLISHER={Freeman Cooper}, 
    ADDRESS={San Francisco},
        PAGES={82-115},
    YEAR={1972},
}

\bib{TGer2008a}{article}{
    AUTHOR={T. Gernhard},
    TITLE={The conditioned reconstructed process},
    JOURNAL={J. Theor. Biol.},
        VOLUME={253},
        PAGES={769-778},
    YEAR={2008},
}

\bib{TGer2008b}{article}{
    AUTHOR={T. Gernhard},
    TITLE={New Analytic Results for Speciation Times in Neutral Models},
    JOURNAL={B. Math. Biol.},
        VOLUME={70},
        PAGES={1082-1097},
    YEAR={2008},
}

\bib{SGouNEld1977}{article}{
    AUTHOR={S. J. Gould},
    author={N. Eldredge},
    TITLE={Punctuated equilibria: the tempo and mode of evolution reconsidered},
    JOURNAL={Paleobiology},
        VOLUME={3},
        NUMBER={2},
        PAGES={115-151},
    YEAR={1977},
}

\bib{SGouNEld1993}{article}{
    AUTHOR={S. J. Gould},
    author={N. Eldredge},
    TITLE={Punctuated equilibrium comes of age},
    JOURNAL={Nature},
        VOLUME={366},
        PAGES={223-227},
    YEAR={1993},
}

\bib{THan1997}{article}{
    AUTHOR={T. F. Hansen},
    TITLE={Stabilizing selection and the comparative analysis of adaptation},
    JOURNAL={Evolution},
        VOLUME={51},
        NUMBER={5},
        PAGES={1341-1351},
    YEAR={1997},
}

\bib{MLanJSchMLia2013}{article}{
    AUTHOR={M. J. Landis},
    author={J. G. Schraiber},
    author={M. Liang},
    TITLE={Phylogenetic analysis using {L}\'evy processes: finding jumps in the evolution of continuous traits},
    JOURNAL={Syst. Biol.},
        VOLUME={62},
        NUMBER={2},
        PAGES={193-204},
    YEAR={2013},
}

\bib{EMay1982}{article}{
    author={E. Mayr},
    TITLE={Speciation and macroevolution},
    JOURNAL={Evolution},
        VOLUME={36},
        PAGES={1119-1132},
    YEAR={1982},
}

\bib{KPet1983}{book}{
    TITLE={Ergodic Theory},
    AUTHOR={K Petersen},
    PUBLISHER={Cambridge University Press},
    ADDRESS={Cambridge},
    YEAR={1983},
}

\bib{SSagKBar2012}{article}{
    author = {S. Sagitov}, 
    author={K. Bartoszek},
    title = {Interspecies correlation for neutrally evolving traits},
    journal = {J. Theor. Biol.},
        VOLUME={309},
        PAGES={11-19},
    year = {2012},
}

\bib{TreeSim1}{article}{
    AUTHOR={T. Stadler},
    TITLE={On incomplete sampling under birth-death models and connections to the sampling-based coalescent},
    JOURNAL={J. Theor. Biol.},
        VOLUME={261},
        NUMBER={1},
        PAGES={58-68},
    YEAR={2009},
}

\bib{TreeSim2}{article}{
    AUTHOR={T. Stadler},
    TITLE={Simulating Trees with a Fixed Number of Extant Species},
    JOURNAL={Syst. Biol.},
        VOLUME={60},
        NUMBER={5},
        PAGES={676-684},
    YEAR={2011},
}

%\bib{DWriJHer2011}{article}{
%    AUTHOR={D. B. Wright},
%    author={J. A. Herrington},
%    TITLE={Problematic standard errors and confidence intervals for skewness and kurtosis},
%    JOURNAL={Behav. Res,},
%        VOLUME={43},
%        PAGES={8-17},
%    YEAR={2011},
%}

 \end{bibliography}

\end{document}